\UseRawInputEncoding
\documentclass[%
reprint,
superscriptaddress,
nofootinbib,
 amsmath,amssymb,
 aps,
]{revtex4-1}
\usepackage{graphicx}
\usepackage{subfigure}
\usepackage{dcolumn}
\usepackage{bm}
\usepackage[%
bookmarksnumbered,
bookmarksopen,
colorlinks,
citecolor=blue,
linkcolor=blue,
]{hyperref} 

\def\esym{$E_{sym}(\rho)$~}

\def\es0{$E_{sym}(\rho_0)$~}
\def\us0{$U_{sym}^{\infty}(\rho_{0})$}
\begin{document}


\title{Directed and elliptic flows of protons and deuterons in HADES Au+Au collisions at $\sqrt{s_{\rm NN}}=2.4$~GeV}

\author{Huan Du}
\affiliation{School of Physics and Electronic Science, Guizhou Normal University, Guiyang 550025, China}
\author{Gao-Feng Wei}\email{wei.gaofeng@gznu.edu.cn}
\affiliation{School of Physics and Electronic Science, Guizhou Normal University, Guiyang 550025, China}
\affiliation{Guizhou Provincial Key Laboratory of Radio Astronomy and Data Processing, Guizhou Normal University, Guiyang 550025, China}
\author{Gao-Chan Yong}\email{yonggaochan@impcas.ac.cn}
\affiliation{Institute of Modern Physics, Chinese Academy of Sciences, Lanzhou 730000, China}
\affiliation{School of Nuclear science and Technology, University of Chinese Academy of Sciences, Beijing 100049, China}


\begin{abstract}

Within a transport model coupled with a microscopic coalescence model, the directed and elliptic flows of protons and deuterons as well as their scalling properties are studied in the centrality of 20-30\% Au+Au collisions at $\sqrt{s_{\rm NN}}=2.4$~GeV. It is found that the flows as well as their scaling properties simulated with the isospin- and momentum-dependent nuclear mean field with an incompressibility $K_{0}=230$ MeV fit fairly the HADES data, while those simulated with the commonly used momentum-independent nuclear mean field with an incompressibility $K_{0}=380$ MeV can only fit partially the HADES data. Moreover, by checking the rapidity distributions of both protons and deuterons in the centrality of 0-10\% Au+Au collisions at $\sqrt{s_{\rm NN}}=2.4$~GeV, we find that the rapidity distributions of deuterons are underestimated while those of protons are overestimated by the simulations with the momentum-independent nuclear mean field. In contrast,
the rapidity distributions  of both protons and deuterons simulated with the isospin- and momentum-dependent nuclear mean field are in good agreement with the HADES data. Our findings imply that the momentum dependence of nuclear mean field is an unavoidable feature for a fundamental understanding of nuclear matter properties and for the successful interpretation of the HADES data.

\end{abstract}

\maketitle


\section{introduction}\label{introduction}

Heavy-ion collisions (HICs) can directly generate the high density and/or temperature strong interacting matter, and thus provide the opportunity to explore the strong interaction properties at extreme conditions.
One of the interests in HIC community is the exploration of nuclear equation of state (EoS) as well as the symmetry energy for asymmetric nuclear matter at high densities due to its crucial role in understanding not only  the compact stellar objects, e.g., neutron stars~\cite{Latt16,Oert17,Lim18,Tews18} but also the later evolution of the high temperature quark-gluon-plasma (QGP) created in HICs as well as the associated QCD phase transition~\cite{Gupta11,Braun15,Munz16}.

The collective motion of final state particles is a direct reflection of the pressure and its gradient created in HICs and thus is closely related to the EoS of dense matter. The directed flow and the elliptic flow are the first two Fourier coefficients of the azimuthal distribution of the final state particles in the momentum space~\cite{Posk98}, and are widely used as observables to extract information of the EoS from HICs~\cite{Sturm01,Fuch01,Dan02,Hart06,Pan93,Zhang94,Stock86}. With these pioneer efforts~\cite{Sturm01,Fuch01,Dan02,Hart06,Pan93,Zhang94,Stock86}, it seems that two understandings related to the EoS are achieved. First, it is most likely that the momentum-dependent nuclear mean field with an incompressibility of $K_{0}=230\pm30$ MeV could describe experimental flow data of HICs within a broad range of collision energies. Second, it seems that the momentum-independent nuclear mean field with an incompressibility of approximate $K_{0}=380$ MeV could also give good description of the flow data of HICs. Interestingly, it is noticed very recently that these two views are well supported by two recent studies~\cite{Hillm20,fang22}, respectively. Specifically, within the ultra relativistic quantum molecular dynamics (UrQMD) model, the authors of Ref.~\cite{Hillm20} claimed that the HADES flow data could be well fitted by a momentum-independent nuclear mean field with an incompressibility of approximate $K_{0}=380$ MeV, while in Ref.~\cite{fang22} the HADES flow data are well reproduced by a momentum-dependent nuclear mean field within the framework of isospin dependent quantum molecular dynamics (IQMD) model. To this situation, a natural question is whether these two interactions produce the same mean field effects on the directed and/or elliptic flows. Moreover,  whether the Boltzmann-Uehling-Uhlenbeck (BUU) model could fit the flow data well as the QMD-type models with these two kinds of nuclear mean fields. In addition, one may want to know whether the directed and/or elliptic flows in HADES Au + Au collisions could be used to probe the symmetry energy at high densities, especially above twice
saturation density. This naturally calls for a comparative study with these two kinds of nuclear mean field within a BUU transport model. To this end, within a BUU transport model coupled with a microscopic coalescence model, the directed and elliptic flows of protons and deuterons as well as their scalling properties are studied with two nuclear mean field scenarios in the HADES Au+Au collisions at $\sqrt{s_{\rm NN}}=2.4$~GeV~\cite{HADES,ECT}. It is found that the momentum dependence of nuclear mean field plays an essential role for the successful interpretation of the HADES data, while the symmetry energy effect seems to be negligible.
\section{The Model}\label{Model}
The current study is carried out within an isospin- and momentum-dependent BUU (IBUU) transport model~\cite{Das03,IBUU,Chen05}. To perform the comparative study as aforementioned, we  incorporate two nuclear mean field scenarios in this IBUU model as briefly described in the following.

The first is a commonly used momentum-independent nuclear mean field with an incompressibility of approximate $K_{0}=380$~MeV,
since this potential could also reproduce the HADES flow data as reported, e.g., in Ref.~\cite{Hillm20} as aforementioned. However, to examine the possible isospin effects on the flows simultaneously, we also consider the symmetry energy effects in the simulation with this interaction.
Specifically, the interaction used in this study (labelled as MID here) is expressed as, 
\begin{equation}
U(\rho,\delta,\tau)=\alpha{\Big(}\frac{\rho}{\rho_{0}}{\Big)}+\beta{\Big(}{\frac{\rho}{\rho_{0}}}{\Big)}^{\xi}+v_{asy}(\rho,\delta,\tau),
\end{equation}
where $\alpha=-123.03$ MeV, $\beta=69.77$ MeV, and $\xi=2.01$. The first two terms on the right of the above expression are exactly the commonly used momentum-independent nuclear mean field, e.g., used in Ref.~\cite{Hillm20}, and the values used here for parameters $\alpha$ and $\beta$ lead to an incompressibility of $K_{0}=380$~MeV. The third term, i.e., $v_{asy}(\rho,\delta,\tau)$ expressed as~\cite{LiBA02}, 
\begin{eqnarray}%
v_{asy}(\rho,\delta,\tau)&=&\tau{\Big(}E_{sym}(\rho_{0})u^{\gamma}-12.7u^{2/3}{\Big)}\delta \notag \\
&+&{\Big(}E_{sym}(\rho_{0})(\gamma-1)u^{\gamma}+4.2u^{2/3}{\Big)}\delta^{2},
\end{eqnarray}%
is exactly the isovector part we considered in this study that results in the symmetry energy with a form 
$E_{sym}(\rho)=E_{sym}(\rho_{0})u^{\gamma}$,
where $u\equiv{\rho/\rho_{0}}$ is the reduced density, $\delta$ is the isospin asymmetry, $\tau$ is 1 for neutrons and $-1$ for protons, and $E_{sym}(\rho_{0})=32.5$ MeV is the symmetry energy at the normal density. The parameter $\gamma$ is used to adjust the stiffness of symmetry energy. In this study, we consider the two cases of 􏰬$\gamma=0.5$ 􏰰(soft) and 􏰬$\gamma=2$ (stiff) to explore the large range of symmetry energy effects predicted by many-body theories. The corresponding slope values $L\equiv{3\rho({dE_{sym}}/d\rho})$ of \esym at $\rho_{0}$ for the two cases of $\gamma$ values are listed in Table~\ref{tableI}.
\begin{table}[hbt]
	\caption{\label{tableI}
		The incompressibility $K_{0}$ and the parameter $\gamma$ as well as the corresponding $L$ values for MID interactions.}
	\begin{ruledtabular}
		\begin{tabular}{lccc}
			\textrm{Interactions}&
			\textrm{$K_{0}$ (MeV)}&
			\textrm{$\gamma$}&
			\textrm{$L$ (MeV)}\\
			\colrule
			${\rm MID}$     & $380$ & $0.5$ & $48.75$\\
			${\rm MID}$     & $380$ & $2.0$ & $195.0$\\
		\end{tabular}
	\end{ruledtabular}
\end{table}
\begin{table}[t]
	\caption{\label{tableII}
		The incompressibility $K_{0}$ and the parameters $x$ and $z$ as well as the corresponding $L$ values for MDI interactions.}
	\begin{ruledtabular}
		\begin{tabular}{lcccc}
			\textrm{Interactions}&
			\textrm{$x$}&
			\textrm{$z$ (MeV)}&
			\textrm{$K_{0}$ (MeV)}&
			\textrm{$L$ (MeV)}\\
			\colrule
			${\rm MDI}$    & $0.6$ &~$-1.482$ & $230$ & $33.06$\\
			${\rm MDI}$    &$-0.2$ &$2.844$ & $230$ & $92.66$\\
		\end{tabular}
	\end{ruledtabular}
\end{table}

Another is an isospin- and momentum-dependent nuclear mean field (labelled as MDI here) that is updated as using a separate density-dependent scenario for the in-medium nucleon-nucleon interaction~\cite{Wei22}, for the purpose of more delicate treatment of the in-medium many-body force effects~\cite{Chen14,Xu10} as well as studying the isospin physics. Specifically, the MDI interaction used here is expressed as:
\begin{eqnarray}
U(\rho,\delta ,\vec{p},\tau ) &=&A_{u}\frac{\rho _{-\tau }}{\rho _{0}}%
+A_{l}\frac{\rho _{\tau }}{\rho _{0}}+\frac{B}{2}{\Big(}\frac{2\rho_{\tau} }{\rho _{0}}{\Big)}^{\sigma }(1-x)  \notag \\
&+&\frac{2B}{%
	\sigma +1}{\Big(}\frac{\rho}{\rho _{0}}{\Big)}^{\sigma }(1+x)\frac{\rho_{-\tau}}{\rho}{\big[}1+(\sigma-1)\frac{\rho_{\tau}}{\rho}{\big]}
\notag \\
&+&\frac{2C_{l }}{\rho _{0}}\int d^{3}p^{\prime }\frac{f_{\tau }(%
	\vec{p}^{\prime })}{1+(\vec{p}-\vec{p}^{\prime })^{2}/\Lambda ^{2}}
\notag \\
&+&\frac{2C_{u }}{\rho _{0}}\int d^{3}p^{\prime }\frac{f_{-\tau }(%
	\vec{p}^{\prime })}{1+(\vec{p}-\vec{p}^{\prime })^{2}/\Lambda ^{2}},
\label{IMDIU}
\end{eqnarray}%
where $\tau=1$ for neutrons and $-1$ for protons, and $A_{u}$, $A_{l}$, $C_{u}(\equiv C_{\tau,-\tau})$ and $C_{l}(\equiv C_{\tau,\tau})$ are expressed as
\begin{eqnarray*}
A_{l}&=&A_{l0}+U_{sym}^{\infty}(\rho_{0}) - \frac{2B}{\sigma+1}\notag \\
&\times&\Big{[}\frac{(1-x)}{4}\sigma(\sigma+1)-\frac{1+x}{2}\Big{]},  \\
A_{u}&=&A_{u0}-U_{sym}^{\infty}(\rho_{0}) + \frac{2B}{\sigma+1}\notag \\
&\times&\Big{[}\frac{(1-x)}{4}\sigma(\sigma+1)-\frac{1+x}{2}\Big{]},\\
C_{l}&=&C_{l0}-2\big{(}U_{sym}^{\infty}(\rho_{0})-2z\big{)}\frac{p_{f0}^{2}}{\Lambda^{2}\ln \big{[}(4p_{f0}^{2}+\Lambda^{2})/\Lambda^{2}\big{]}},\\
C_{u}&=&C_{u0}+2\big{(}U_{sym}^{\infty}(\rho_{0})-2z\big{)}\frac{p_{f0}^{2}}{\Lambda^{2}\ln \big{[}(4p_{f0}^{2}+\Lambda^{2})/\Lambda^{2}\big{]}}.
\end{eqnarray*}
The eight parameters embedded in above expressions, i.e., $A_{l0}$, $A_{u0}$, $B$, $\sigma$, $C_{l0}$, $C_{u0}$, $\Lambda$ and $z$, are determined by fitting experimental/empirical constraints on properties of nuclear matter at $\rho_{0}=0.16$~fm$^{-3}$, i.e., the binding energy $-16$~MeV, the pressure $P_{0}=0$~MeV/fm$^{3}$, the incompressibility $K_{0}=230$~MeV for symmetry nuclear matter (SNM), the isoscalar effective mass $m^{*}_{s}=0.7m$, the isoscalar potential at infinitely large nucleon momentum $U^{\infty}_{0}(\rho_{0})=75$~MeV, the isovector potential at infinitely large nucleon momentum $U^{\infty}_{sym}(\rho_{0})=-100$~MeV as well as the symmetry energy $E_{sym}(\rho)$ at both $\rho_{0}$ and $2\rho_{0}/3$, i.e., $E_{sym}(\rho_{0})=32.5\pm{3.2}$~MeV~\cite{Wang18} and $E_{sym}(2\rho_{0}/3)=25.5\pm{1}$~MeV~\cite{Wang13,Brown13,Dan14,Cozma18,Stone17}. Specifically, the values of first seven parameters are $A_{l0}=A_{u0}=-66.963$~MeV, $B=141.963$~MeV, $C_{l0}=-60.486$~MeV, $C_{u0}=-99.702$~MeV, $\sigma=1.2652$, and $\Lambda=2.424p_{f0}$, where $p_{f0}$ is the nucleon Fermi momentum in SNM at $\rho_{0}$. The eighth parameter $z$ is used to mimic the value of $E_{sym}(\rho_{0})$ in the allowed range to ensure the value of $E_{sym}(2\rho_{0}/3)=25.5\pm1$~MeV since the best knowledge of \esym we can obtain  so far is around $2\rho_{0}/3$~\cite{Wang13,Brown13,Dan14,Cozma18,Stone17}.
Moreover, the parameter $x$ is used to mimic the slope value $L$ of \esym at $\rho_{0}$ without changing the value of $E_{sym}(\rho)$ at $2\rho_{0}/3$ and any property of the SNM. The values of parameters $x$ and $z$ used in this study as well as the corresponding $L$ values are listed in Table~\ref{tableII}.

The coalescence model has been used extensively in studying the production of light clusters in HICs at both intermediate~\cite{Gyul83,Aich87,Koch90} and relativistic~\cite{Nagl96,Matt97,Sche99,Poll99} energies. In this study, we use a microscopic coalescence model~\cite{Chen03a,Chen03b} to study the production of light nuclei. In this model, the probability for producing a cluster is determined by the overlap of nucleon Wigner phase-space density with the nucleon phase-space distributions at the freeze-out. Specifically, the  multiplicity of a $M$-nucleon cluster is determined by~\cite{Matt97}
\begin{eqnarray}
N_{M}&=&G\int \sum_{i_{1}>{i_{2}>\cdot\cdot\cdot>{i_{M}}}} d{{\bf r}_{i_{1}}}d{{\bf k}_{i_{1}}}\cdot\cdot\cdot d{{\bf r}_{i_{M-1}}}d{{\bf k}_{i_{M-1}}}\notag \\
&\times&\left \langle \rho_{i}^{W}({{\bf r}_{i_{1}}},{{\bf k}_{i_{1}}},\cdot\cdot\cdot, {{\bf r}_{i_{M-1}}},{{\bf k}_{i_{M-1}}}) \right \rangle,
\end{eqnarray}
where ${\bf r}_{i_{1}}$,$\cdot\cdot\cdot$,${\bf r}_{i_{M-1}}$ and ${\bf k}_{i_{1}}$,$\cdot\cdot\cdot$,${\bf k}_{i_{M-1}}$ are, respectively, the $M-1$ relative coordinates and momenta in the $M$-nucleon rest frame, $\left \langle\cdot\cdot\cdot\right\rangle$ denotes the event averaging, and $\rho_{i}^{W}$ is the Wigner phase space density of the $M$-nucleon cluster.
More specifically, for the deuteron considered in this study, the Wigner phase space density is obtained from the Hulth$\acute{e}$n wave function
with adjusting to reproduce the measured deuteron root-mean-square radius of 1.96 fm.
The spin-isospin statistical factor for the deuteron is given by $G$ with the value of $3/8$. For more details of this coalescence model, see, e.g., Refs.~\cite{Matt97,Chen03b}.
\begin{figure}[thb]
	\includegraphics[width=\columnwidth]{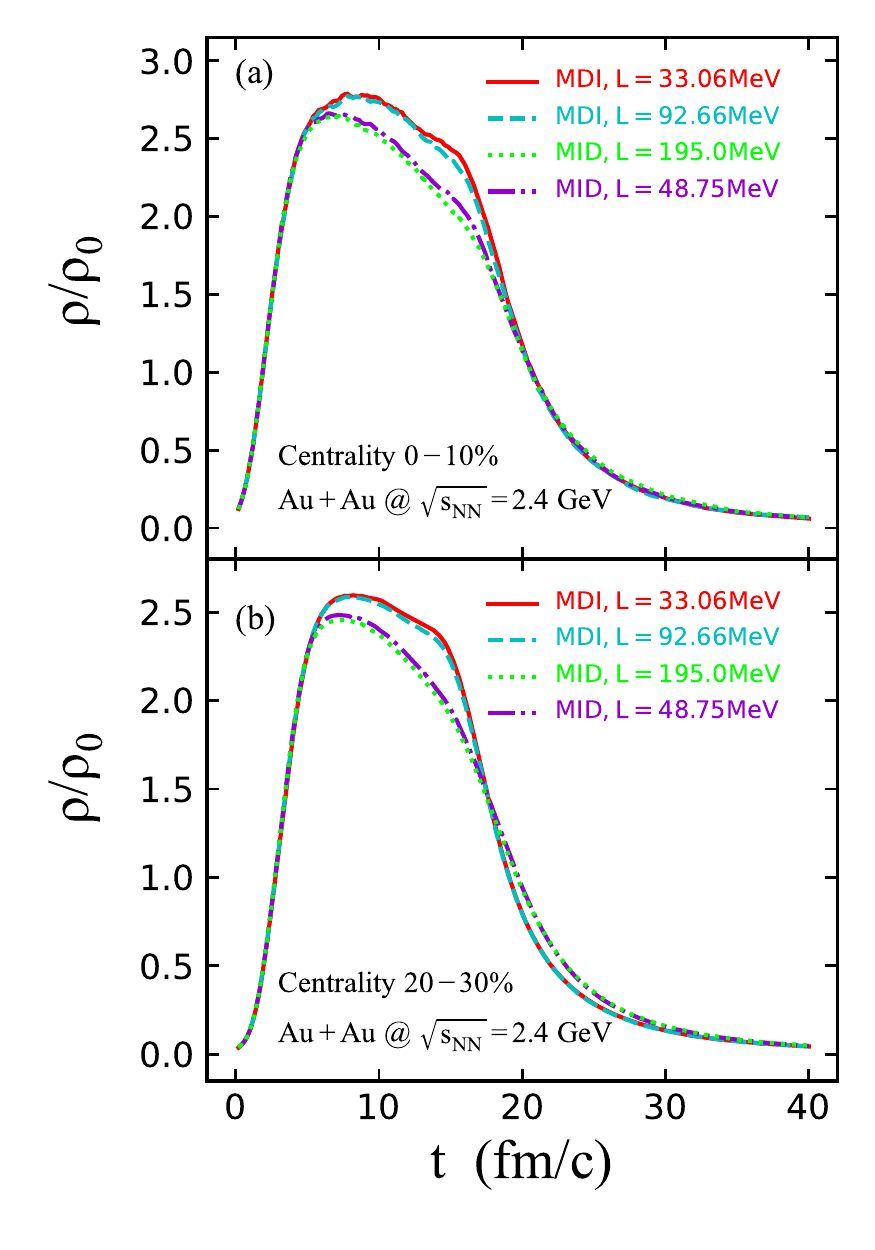}
	\caption{(Color online) Evolution of central region reduced density $\rho/\rho_{0}$ in 0-10\% (a) and 20-30\% (b) Au+Au collisions at $\sqrt{s_{\rm NN}}=2.4$~GeV.} \label{den}
\end{figure}
\begin{figure*}
	\centering
	\resizebox{0.8\textwidth}{!}{
		\includegraphics{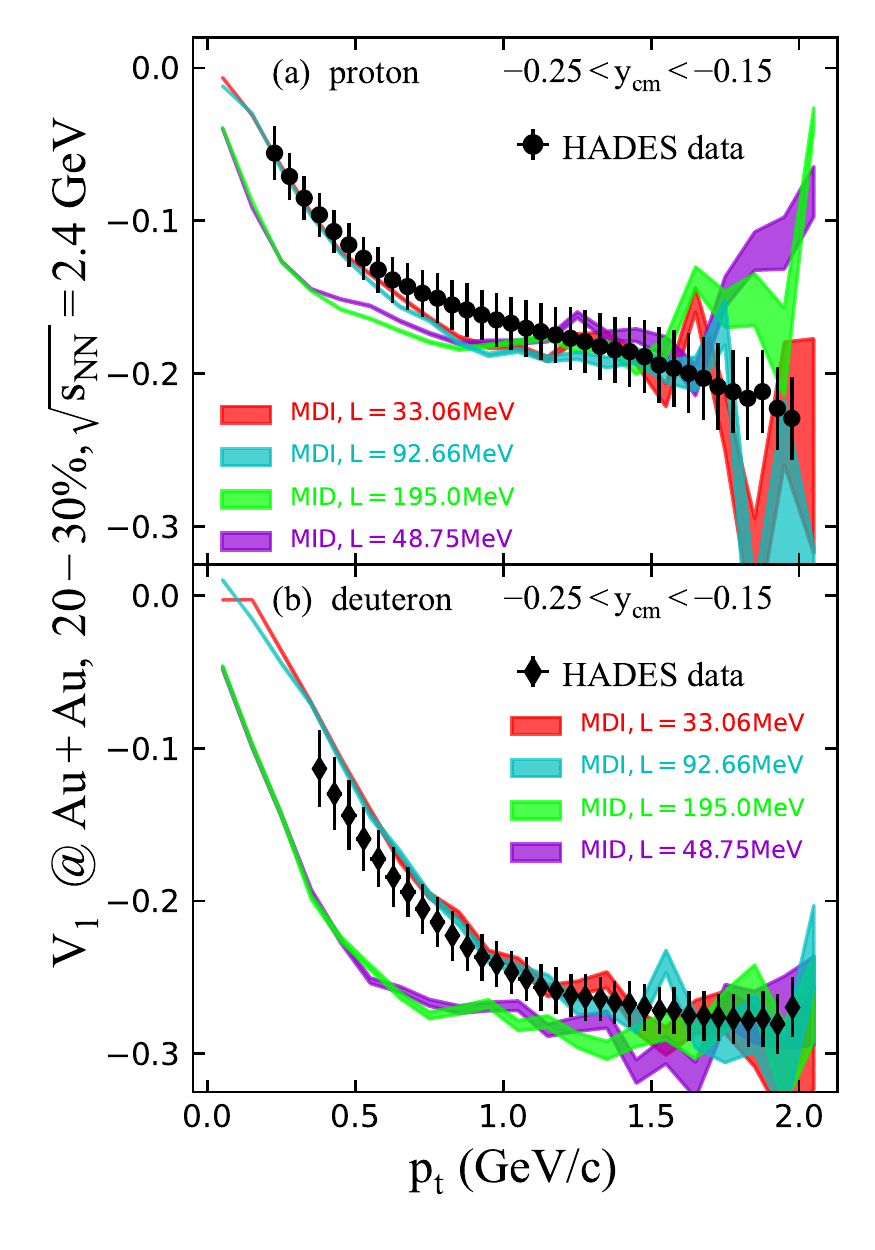}
		\includegraphics{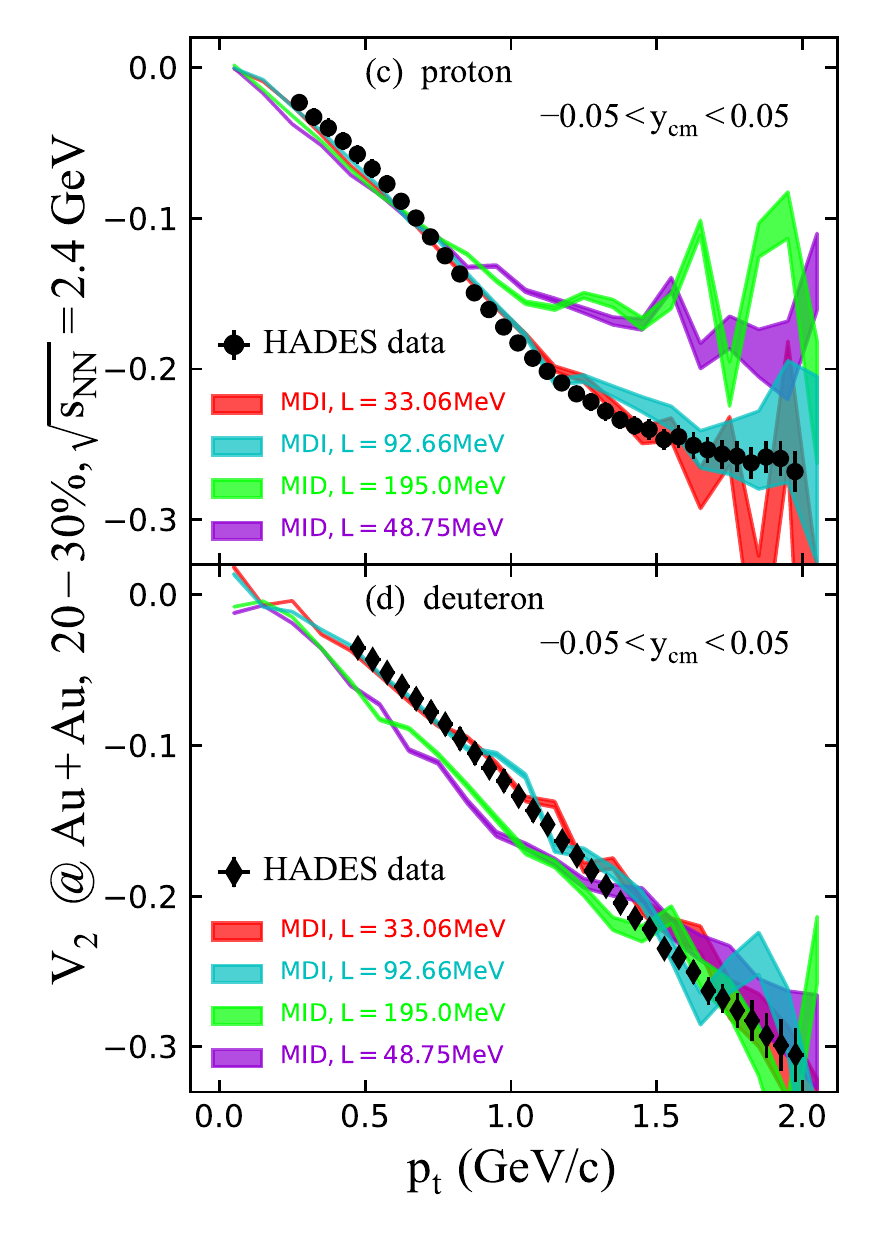}
	}
	\caption{(Color online) The directed and elliptic flows of protons and deuterons as a function of transverse momentum in 20-30\% Au+Au collisions at $\sqrt{s_{\rm NN}}=2.4$~GeV in comparison with the HADES data.}\label{flow-pt}
\end{figure*}
\begin{figure*}
	\centering
	\resizebox{0.8\textwidth}{!}{
		\includegraphics{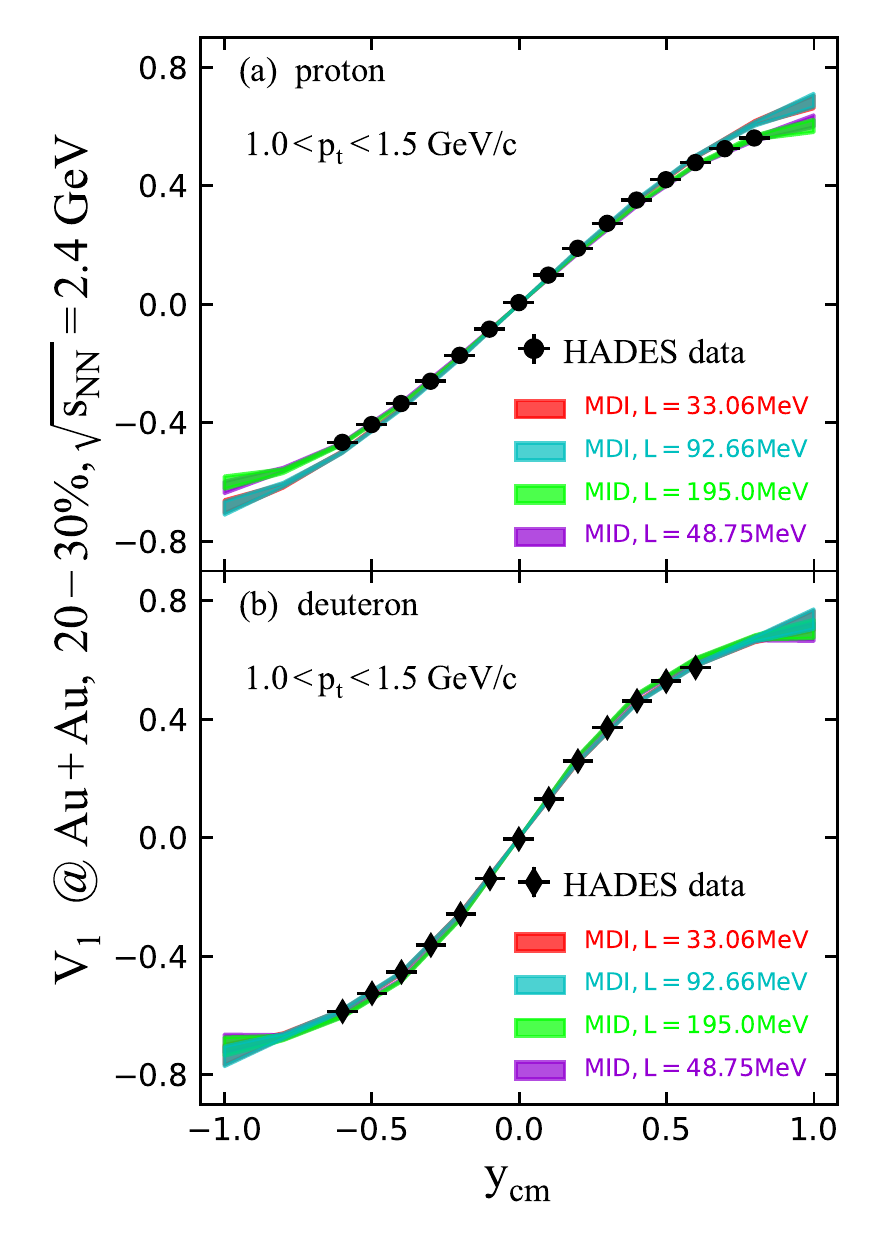}
		\includegraphics{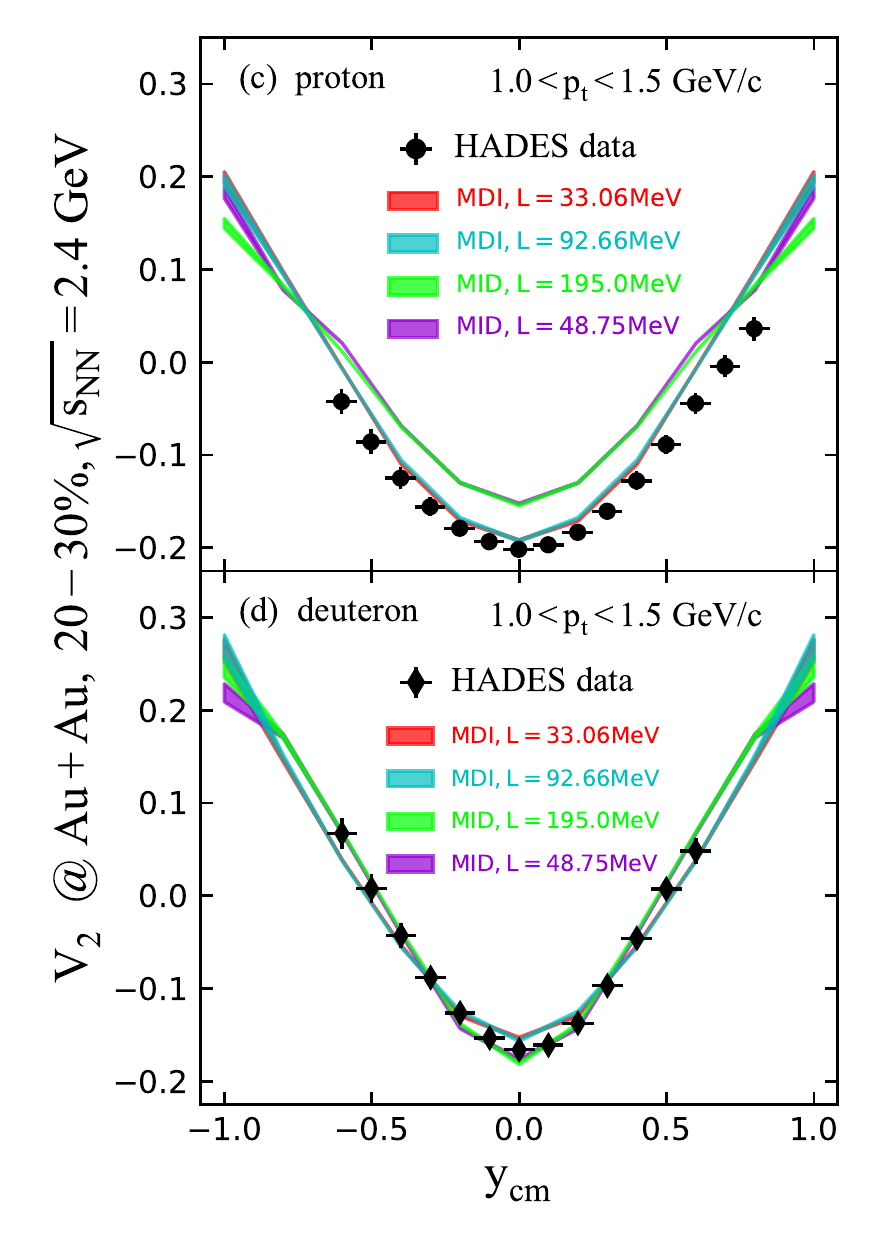}
	}
	\caption{(Color online) The directed and elliptic flows of protons and deuterons as a function of rapidity in center of mass frame in 20-30\% Au+Au collisions at $\sqrt{s_{\rm NN}}=2.4$~GeV in comparison with the HADES data.}\label{flow-rap}
\end{figure*}

\section{Results and Discussions}\label{Results and Discussions}

Now, we present results of Au+Au collisions at $\sqrt{s_{\rm NN}}=2.4$~GeV. To help understand the directed and elliptic flows simulated with the two nuclear mean field scenarios as aforementioned, it is meaningful to show the compression density of central region reached in the studied reactions. Figure~\ref{den} shows the evolution of compression density in the central region reached in Au+Au collisions for two different centralities. One can see that the densities reached in the simulation with the MDI nuclear mean field with an incompressibility $K_{0}=230$ MeV are  significantly larger than those simulated with the MID nuclear mean field with an incompressibility $K_{0}=380$ MeV, reflecting the fact that the nuclear compression is overall sensitive to the bulk EoS of nuclear matter, and the symmetry energy effect is decreased to a negligible degree as the beam energy increases up to approximate 1 GeV and above. Interestingly, in the more non-central collisions, e.g.,  the Au + Au collisions with the centrality of 20-30\% studied here, we observe that the compression density reached with the MDI mean field scenario is becoming smaller after approximate 18 fm/c than that with the MID mean field scenario. This is because the MDI nuclear mean field gets the compression region to form more elliptic shape as well as larger squeeze-out pressure that causes more emissions of the high transverse momentum nucleons in the compression stage, and thus leads to a smaller density in the central region in the later expansion stage. These observations naturally are expected to be reflected by the collective motion of emitting particles, especially the high transverse momentum squeeze-out particles emitted from the mid-rapidities.

Shown in Fig.~\ref{flow-pt} are the transverse momentum dependent directed and elliptic flows of protons and deuterons in 20-30\% Au+Au collisions at $\sqrt{s_{\rm NN}}=2.4$~GeV with two nuclear mean field scenarios in comparison with the corresponding HADES data. As expected, within the very central rapidity region, e.g., $|{\rm y_{cm}}|<0.05$ as shown in Fig.~\ref{flow-pt}(c), the proton elliptic flow at the high transverse momenta is indeed significantly more negative with the MDI nuclear mean field scenario than that with the MID nuclear mean field scenario\footnote{Throughout this study, when comparing the observables, we consider them along with their numerical signs.}, reflecting the stronger squeeze-out effects of the MDI nuclear mean field. And even, in the projectile rapidity region of $-0.25<{\rm y_{cm}}<-0.15$ as shown in Fig.~\ref{flow-pt}(a), this regularity also holds for the directed flow of protons at the transverse momentum larger than 1.5 GeV/c. In contrast, for the later emission in the reaction expansion stage, since the pressure generated by the MDI nuclear mean field is smaller than that by the MID nuclear mean field,  it is therefore the proton directed flow in the MID case has a more negative value at the lower transverse momenta. Clearly, it is seen that the simulation with the MDI nuclear mean field could fit fairly both the proton directed and elliptic flows of the HADES data, while the elliptic (directed) flow of protons at the higher (lower) transverse momenta is underestimated (overestimated) with the MID nuclear mean field\footnote{Here, underestimate/overestimate means that the absolute value of the simulated flow is smaller/larger than that of the corresponding experimental measurement value.}. In addition, we still can observe a weak symmetry energy effects from the proton directed and elliptic flows at the very high transverse  momenta, e.g., around 2 GeV/c, the elliptic flow of protons with a soft symmetry energy (a small $L$ value) is slightly smaller than that with a stiff one (a large $L$ value) due to the slightly larger compression generated by the soft symmetry energy as shown in Fig.~\ref{den}. These observations are in fact the direct reflection of the pressure and its gradient created in non-central collisions. However, unlike the proton directed and elliptic flows, the effects of nuclear mean field on the deuteron directed and elliptic flows are mainly located at the lower transverse momenta, in particular for the deuteron directed flows as shown in Fig.~\ref{flow-pt}(b). This is exactly due to the deuteron is formed at the relatively lower densities, both its directed and elliptic flows are affected mainly by the pressure created in the expansion stage. Therefore, it is not hard to understand the directed and elliptic flows of deuterons are both smaller with the MID nuclear mean field scenario than those with the MDI nuclear mean field scenario. Moreover, due to the deuteron is composed of a neutron and a proton, and its formation is usually in the expansion stage, therefore both its directed and elliptic flows are insensitive to the symmetry energy.

\begin{figure}[thb]
	\includegraphics[width=\columnwidth]{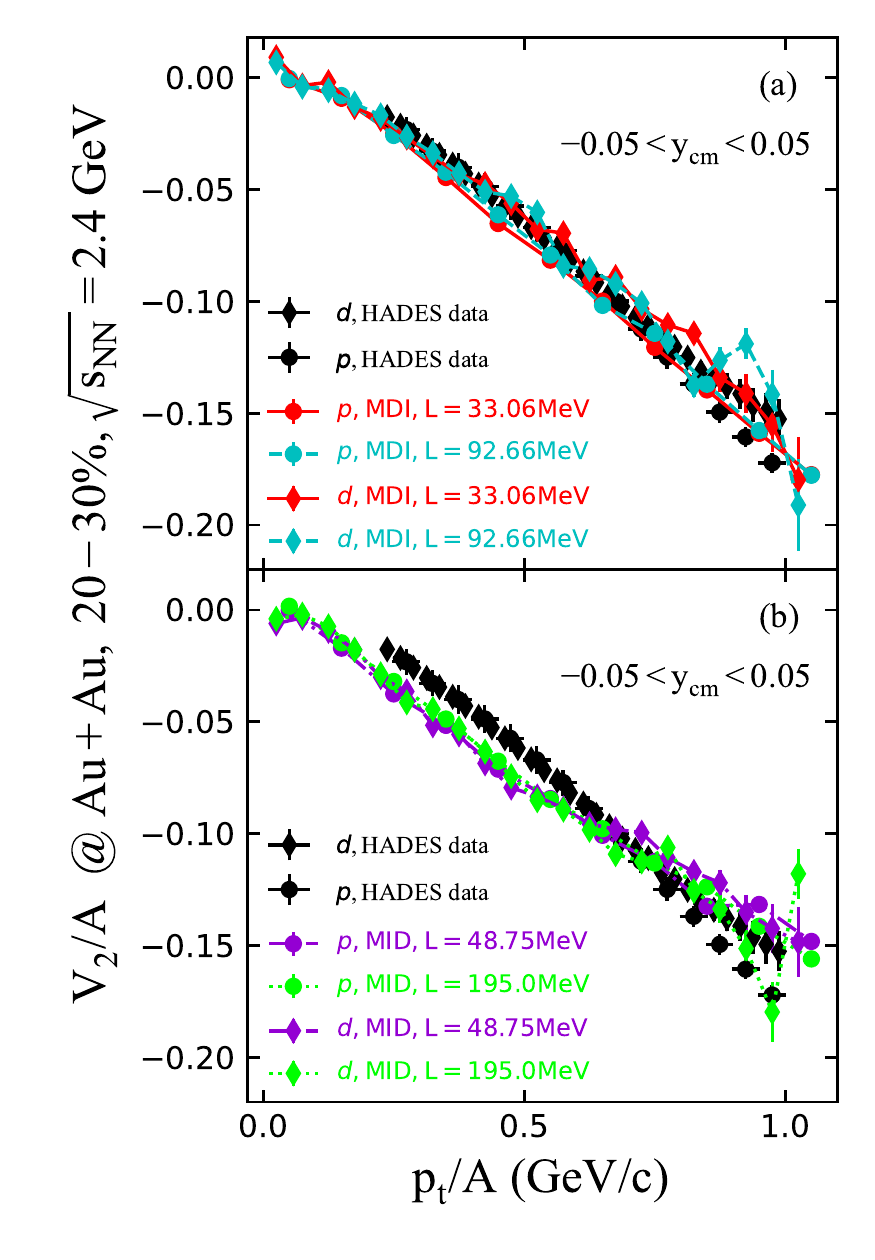}
	\caption{(Color online)The nucleon-number scaled elliptic flows of protons and deuterons as a function of transverse momentum per nucleon in 20-30\% Au+Au collisions at $\sqrt{s_{\rm NN}}=2.4$~GeV. }
	\label{scaling-ptA}
\end{figure}
\begin{figure}[hbt]
	\includegraphics[width=\columnwidth]{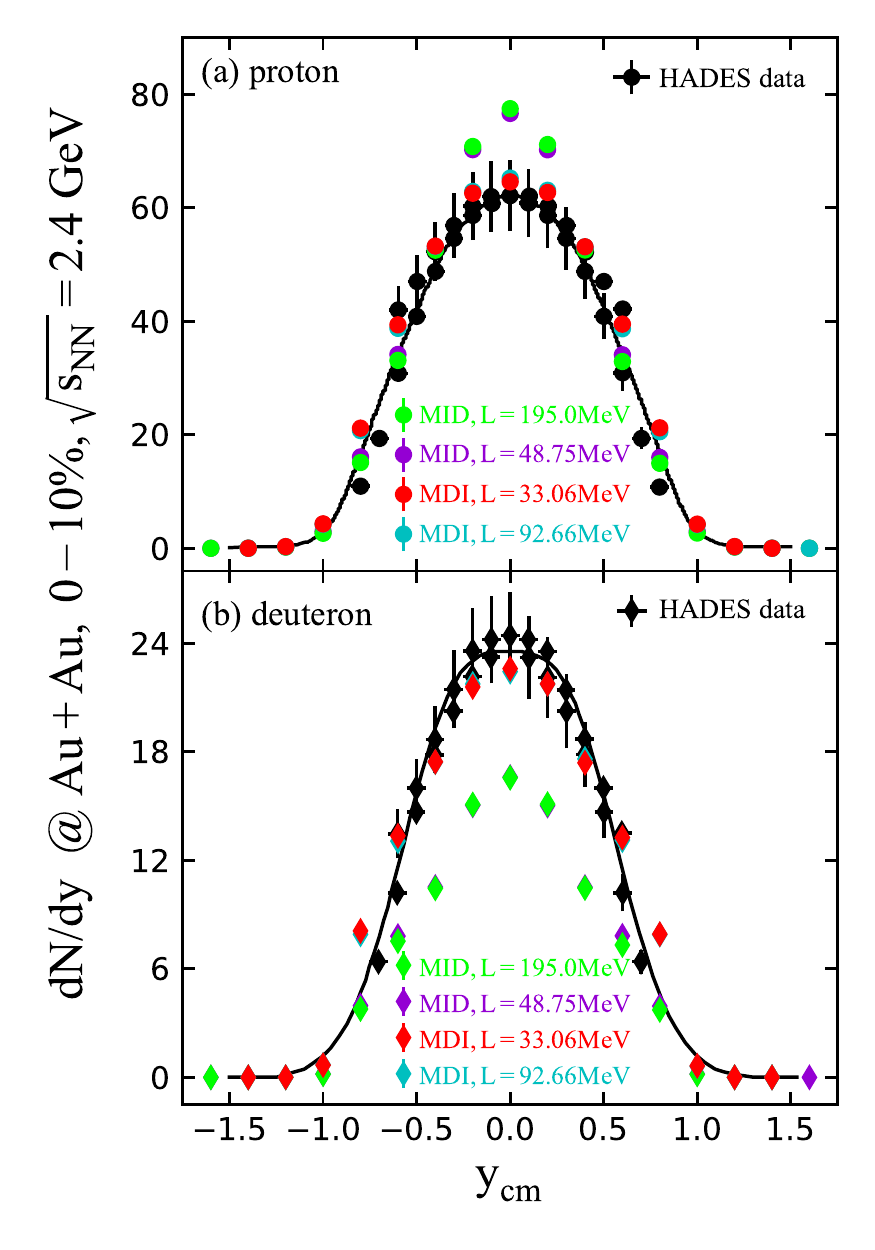}
	\caption{(Color online) Rapidity distribution of protons (a) and deuterons (b) in 0-10\% Au+Au collisions at $\sqrt{s_{\rm NN}}=2.4$~GeV in comparison with the HADES data. }
	\label{rap-dis}
\end{figure}

Shown in Fig.~\ref{flow-rap} are the rapidity dependent directed and elliptic flows of protons and deuterons in 20-30\% Au+Au collisions at $\sqrt{s_{\rm NN}}=2.4$~GeV with two nuclear mean field scenarios in comparison with the corresponding HADES data. Interestingly, except for the proton elliptic flow, it is seen that the simulations with both the MDI and MID mean field scenarios can fit the HADES data well. While for the proton elliptic flow as shown in Fig.~\ref{flow-rap}(c), we observe again a more negative value at the mid-rapidity region in the MDI mean field scenario, reflecting a more strong squeeze-out effect generated by the MDI mean field. Certainly, we still can observe a very weak effect of nuclear mean field on the proton directed flow and deuteron elliptic flow as shown in Fig.~\ref{flow-rap}(a) and Fig.~\ref{flow-rap}(d). Actually, it is not surprising to see only a mild difference between the MDI and MID scenarios for the directed flows, since the data are only for the large transverse momentum range from 1 to 1.5 GeV/c.

It is well known that the constituent-number scaling of elliptic flows of hadrons is an inherent characteristic of coalescence mechanism of hadrons produced in HICs~\cite{Frie03,Moln03,Kolh04,Yan06,Adam16,Abda22}. Therefore, studying the nucleon-number scaling of elliptic flows of deuterons is useful to further verify the coalescence mechanism used in this study. Similar to the number of constituent quark scaling of elliptic flows of hadrons~\cite{Frie03,Moln03,Kolh04}, the deuteron elliptic flow is expected to follow an approximate mass number $A$ scaling~\cite{Yan06,Adam16,Abda22}, i.e.,
\begin{equation}
v_{2}^{d}(p_{t},y)/A\approx v_{2}^{p}(p_{t}/A,y).
\end{equation}
Therefore, according to this scaling regularity, the $v_{2}/A$ as a function of $p_{t}/A$ yields approximately the same curves for protons and deuterons. Shown in Fig.~\ref{scaling-ptA}(a) and Fig.~\ref{scaling-ptA}(b) are the corresponding results from the MDI and MID mean field scenarios in comparison with the corresponding HADES data. Clearly, it is seen that the results with the MDI mean field scenario are in good agreement with the data, and thus capture the scaling regularity of coalescence mechanism for deuterons, while those with the MID mean field scenario deviate significantly from the data. More specifically, the scaled elliptic flow is underestimated at the high transverse momenta while that at the low transverse momenta is overestimated, consistent with the previous observations as shown in Fig.~\ref{flow-pt}. These results imply that the momentum dependence of nuclear mean field plays an essential role for the successful interpretation of the HADES data. Before ending this part, we compare again in Fig.~\ref{rap-dis} the rapidity distribution of protons and deuterons in 0-10\% Au + Au collisions at $\sqrt{s_{\rm NN}}=2.4$~GeV with the two mean field scenarios to further show the momentum dependence of nuclear mean field. It is seen again that the rapidity distributions of both protons and deuterons with the MDI mean field scenario are in good agreement with the data, and due to lack of the momentum dependence in the MID nuclear mean field, the rapidity distribution of deuterons is underestimated while that of protons is overestimated regardless of the symmetry energy parameter $\gamma$ is used. The significant enhancement of the deuteron yields with the MDI mean field scenario compared to the MID mean field scenario is consistent with the previous study~\cite{Chen04}.
From these findings, we therefore could conclude that the momentum dependence of nuclear mean field is an unavoidable feature for a fundamental understanding of nuclear matter properties and for the successful interpretation of the HADES data.

\section{Summary}\label{Summary}

Exploration of the EoS as well as the symmetry energy of asymmetric nuclear matter at high densities is a long-standing question in both nuclear physics and astrophysics. The recent HADES experiments for the Au+Au collisions at $\sqrt{s_{\rm NN}}=2.4$~GeV produce dense matter with densities of about 2.5 times saturation density, thus provide a good opportunity to constrain the EoS and/or symmetry energy at high densities. In the framework of an isospin- and momentum-dependent transport model coupled with a microscopic coalescence model, the directed and elliptic flows of protons and deuterons as well as their scalling properties are studied in the Au+Au collisions at $\sqrt{s_{\rm NN}}=2.4$~GeV. It is found that the flows as well as their scaling properties simulated with the MDI nuclear mean field with an incompressibility $K_{0}=230$ MeV fit fairly the HADES data, while those simulated with the commonly used MID nuclear mean field with an incompressibility $K_{0}=380$ MeV can only fit partially the HADES data. The findings imply that the momentum dependence of nuclear mean field is an unavoidable feature for a fundamental understanding of nuclear matter properties and for the successful interpretation of the HADES data.

\begin{acknowledgments}
Gao-Feng Wei appreciates Prof. Lie-Wen Chen for his helpful discussion on the coalescence mechanism of the deuteron and his carefully reading the manuscript. 
This work is supported by the National Natural Science Foundation of China under grant Nos.11965008, 12275322, 11405128 and Guizhou Provincial Science and Technology Foundation under Grant No.[2020]1Y034, and the PhD-funded project of Guizhou Normal university (Grant No.GZNUD[2018]11).

\end{acknowledgments}

\end{document}